\documentclass[prd,aps,showpacs,preprintnumbers,amssymb]{revtex4}

\usepackage{color}
\usepackage{epsf}

\def\e3p{$\eta \rightarrow 3 \pi$}

\begin{document}
\title{%
\hfill{\normalsize\vbox{%
\hbox{}
 }}\\
{About the mass corrections in an abelian Higgs model}}

\author{Renata Jora
$^{\it \bf a}$~\footnote[2]{Email:
 rjora@theory.nipne.ro}}

\affiliation{$^{\bf \it a}$ National Institute of Physics and Nuclear Engineering PO Box MG-6, Bucharest-Magurele, Romania}

\date{\today}

\begin{abstract}
We study the corrections to the scalar boson mass in an abelian Higgs model by considering the global properties of the partition function. We find that although the two point function for the Higgs boson receives quantum corrections, the mass of the scalar remains unaltered  and thus the physical mass is equal to the bare one.
\end{abstract}
\maketitle

\section{Introduction}
In the context in which the Higgs boson of the standard model has been at the forefront of both theoretical and experimental search for decades and the corrections to the Higgs boson mass or more exactly the issues associated to them have triggered a flurry of theoretical papers it is important to study the problems associated with scalars and their masses in smaller set-ups or more amenable models. The simple $\Phi^4$ theory has been the subject of many investigations \cite{Wilson}-\cite{Jora1} in the last few decades, especially with regard to the behavior of the renormalized coupling constant $\lambda_R$ in the limit of large cut-off, the so-called "triviality" problem.  Recent studies suggest that indeed the $\Phi^4$ theory is trivial not only in the regime of large bare coupling constant \cite{Frasca2}, \cite{Jora} but also for the full range of values of this \cite{Jora1}. It is important to analyze if the same conclusions can be found in slightly more complicated theories that include scalars.

 In this work we extend the approach employed in \cite{Jora1}  to study the two point correlator and the corrections to the Higgs mass in an abelian Higgs model with spontaneous symmetry breaking. In section II we present the model and also its counterpart in the Fourier space on a lattice. In section III we analyze the corrections to the propagator in the approach introduced in \cite{Jora1}. Section IV is dedicated to conclusions.


\section{The abelian Higgs model}

We consider the abelian Higgs model with the Lagrangian:

\begin{eqnarray}
{\cal L}=-\frac{1}{4}F_{\mu\nu}F^{\mu\nu}+(D^{\mu}\Phi)^{\dagger}(D_{\mu}\Phi)-V(\Phi),
\label{lagr4435}
\end{eqnarray}
where:
\begin{eqnarray}
&&D_{\mu}=\partial_{\mu}+ieA_{\mu}
\nonumber\\
&&V(\Phi)=-\mu^2\Phi^*\Phi+\frac{\lambda}{2}(\Phi^*\Phi)^2.
\label{def443552}
\end{eqnarray}
This model displays spontaneous symmetry breaking at the minimum of the potential:
\begin{eqnarray}
\Phi_0^2=v^2=\left(\frac{\mu^2}{\lambda}\right).
\label{m88}
\end{eqnarray}

We expand the field around its vev to get:
\begin{eqnarray}
\Phi(x)=v+\frac{1}{\sqrt{2}}(h(x)+i\phi(x)),
\label{res6645353}
\end{eqnarray}
where $\phi(x)$ is the Goldstone boson. We shall work in the $R_{\xi}$ gauges with $\xi=1$ where the gauge fixed Lagrangian has the form:
\begin{eqnarray}
&&{\cal L}=-\frac{1}{2}A_{\mu}(-g^{\mu\nu}\partial^2+e^2v^2g^{\mu\nu})A_{\nu}+
\nonumber\\
&&\frac{1}{2}(\partial_{\mu}h)^2-\frac{1}{2}(2\lambda v^2)h^2+\frac{1}{2}(\partial_{\mu}\phi)^2-(ev)^2\phi^2+
\nonumber\\
&&e^2v^2A_{\mu}A_{\nu}g^{\mu\nu}+\frac{2}{\sqrt{2}}e^2vhA_{\mu}A_{\nu}g^{\mu\nu}+\frac{e^2}{2}h^2A_{\mu}A_{\nu}g^{\mu\nu}-
\nonumber\\
&&(-\frac{\mu^4}{2\lambda}+\frac{1}{8}\lambda h^4+\frac{1}{8}\lambda\phi^4+\frac{\lambda}{\sqrt{2}}vh^3+\frac{\lambda}{4}h^2\phi^2+\frac{\lambda}{\sqrt{2}}hv\phi^2)+
\nonumber\\
&&\bar{c}(-\partial^2+2(ev)^2(1+\frac{h}{v}))c.
\label{fulllagr656}
\end{eqnarray}
The last term in Eq. (\ref{fulllagr656}) corresponds to the ghost term of the gauge fixed Lagrangian.

We shall write only the quadratic term of the space time integral of the Lagrangian  ${\cal L}$ on lattice with volume $V$ in the Fourier space:
\begin{eqnarray}
\int d^4 x {\cal L }_2=
&&-\frac{1}{2}\frac{1}{V}\sum_nA_{\mu}(p_n)g^{\mu\nu}(p_n^2-m_A^2)A_{\nu}(-p_n)+
\nonumber\\
&&\frac{1}{2}\frac{1}{V}\sum_nh(p_n)(p_n^2-m_h^2)H(-p_n)+
\nonumber\\
&&\frac{1}{2}\frac{1}{V}\sum_n\phi(p_N)(p_n^2-m_A^2)\phi(-p_n)+
\nonumber\\
&&\frac{1}{V}\sum_n\bar{c}(p_n)(p_n^2-m_A^2)c(-p_n).
\label{res442344}
\end{eqnarray}

\section{Quantum corrections to the scalar propagator}

In \cite{Jora1} we showed that in a simple $\Phi^4$ theory with a real scalar field there is no correction to the all orders two point function and thus to the scalar mass, hence the $\Phi^4$ theory is trivial. We shall apply the method presented there to the abelian Higgs model.
We start with the expression for the two point Higgs scalar function in the path integral formalism:

\begin{eqnarray}
\langle \Omega|Th(x_1)h(x_2)|\Omega\rangle=
\lim_{T\rightarrow \infty(1+i\varepsilon)}
\frac{\int d A_{\mu}(x) d h(x) d \phi(x)d \bar{c}(x) d c(x) h(x_1) h(x_2) \exp[i\int d^4 x{\cal L}]}{\int d A_{\mu}(x) d h(x) d \phi(x)d \bar{c}(x) d c(x) \exp[i\int d^4 x{\cal L}]}.
\label{two88767}
\end{eqnarray}
We first write in the Fourier space:
\begin{eqnarray}
h(x_1)h(x_2)=\frac{1}{V^2}\sum_m \exp[-ik_mx_1]h(k_m)\sum_l \exp[-ik_lx_2]h(k_l),
\label{expr665758}
\end{eqnarray}
and furthermore:
\begin{eqnarray}
&&\langle \Omega|Th(x_1)h(x_2)|\Omega\rangle=
\frac{1}{V^2}\sum_{m,l}\exp[-ik_mx_1-ik_lx_2]\times
\nonumber\\
&&\lim_{T\rightarrow \infty(1+i\varepsilon)}\frac{ \prod_{n,p,r,s,t}\int d A_{\mu}(k_n) d h(k_p) d \phi (k_r) d \bar{c}(k_s) dc (k_t)h(k_m)h(k_l)\exp[i \int d^4 x{\cal L}]}
{ \prod_{n,p,r,s,t}\int d A_{\mu}(k_n) d h(k_p) d \phi (k_r) d \bar{c}(k_s) dc (k_t)\exp[i \int d^4 x{\cal L}]},
\label{te5534}
\end{eqnarray}
where the exponent should be expressed also in the Fourier space.

Next we consider the function:
\begin{eqnarray}
&&I_{x-y}=\frac{1}{V}\sum_m\exp[-ik_m(x_1-x_2)]\langle h(k_m)h(-k_m)\rangle=
\nonumber\\
&&\frac{1}{V}\sum_m\exp[-ik_m(x_1-x_2)]\lim_{T\rightarrow \infty(1+i\varepsilon)}\frac{ \prod_{n,p,r,s,t}\int d A_{\mu}(k_n) d h(k_p) d \phi (k_r) d \bar{c}(k_s) dc (k_t)h(k_m)h(-k_m)\exp[i d^4 x{\cal L}]}
{ \prod_{n,p,r,s,t}\int d A_{\mu}(k_n) d h(k_p) d \phi (k_r) d \bar{c}(k_s) dc (k_t)\exp[i d^4 x{\cal L}]}.
\label{alt554677}
\end{eqnarray}
 The relation between Eq. (\ref{te5534}) and Eq. (\ref{alt554677}) is given by:
 \begin{eqnarray}
&&I_{x-y}=\frac{1}{V}\sum_m\exp[-ik_m(x_1-x_2)]\langle h(k_m)h(-k_m)\rangle=
\nonumber\\
&&\frac{1}{V}\sum_m\exp[-ik_m(x_1-x_2)]\int d^4 z_1\exp[ik_mz_1]\int d^4 z_2 \exp[-ik_mz_2]\langle h(z_1)h(z_2)\rangle=
\nonumber\\
&&\frac{1}{V}\sum_m\int d^4 z_1 d^4 z_2 \exp[-ik_m(x_1-x_2-z_1+z_2)]\langle h(z_1)h(z_2)\rangle=
\nonumber\\
&&\int d^4 z_1 d^4 z_2 \delta(x_1-x_2-z_1+z_2)\langle h(z_1)h(z_2)\rangle=
\nonumber\\
&&\int d^4 z_2 \langle h(z_2+x_1-x_2)h(z_2)\rangle.
\label{expluu76889}
\end{eqnarray}
But according to the definition of the two point function the following relation holds:
\begin{eqnarray}
&&\langle h(z_2+x_1-x_2)h(z_2)\rangle=\int \frac{d^4p}{(2\pi)^4}exp[-ip(x_1-x_2)]\frac{1}{p^2-m^2-M(p^2)}=
\nonumber\\
&&\langle \Omega|Th(x_1)h(x_2)|\Omega\rangle,
\label{res44355}
\end{eqnarray}
where $M(p^2)$ is the all order correction to the two point function in the Fourier space of the scalar $h$.
 From Eqs. (\ref{expluu76889}) and (\ref{res44355}) we then infer on the lattice:
 \begin{eqnarray}
 I_{x_1-x_2}= \int d^4 z_2 \langle \Omega|Th(x_1)h(x_2)|\Omega\rangle=V\int\frac{d^4p}{(2\pi)^4}exp[-ip(x_1-x_2)]\frac{1}{p^2-m^2-M(p^2)},
 \label{res4536}
 \end{eqnarray}
 where $V$ is the volume of space time. Here it is understood that:
 \begin{eqnarray}
 \frac{1}{V}\sum_m \rightarrow \int \frac{d^4p}{(2\pi)^4}
 \label{res53442}
 \end{eqnarray}
 in the continuum limit.

This shows that in order to determine the two point function for the Higgs boson it is sufficient to evaluate $I_{x_1-x_2}$. In order to do that we first separate from the Lagrangian the quadratic part as in Eq.(\ref{res442344}). We then write:
\begin{eqnarray}
&&\langle h(k_m)h(-k_m)\rangle=
\nonumber\\
&&\frac{ -2iV \frac{\delta}{\delta k_m^2}\prod_{n,p,r,s,t}\int d A_{\mu}(k_n) d h(k_p) d \phi (k_r) d \bar{c}(k_s) dc (k_t)\exp[i \int d^4 x{\cal L}]}
{ \prod_{n,p,r,s,t}\int d A_{\mu}(k_n) d h(k_p) d \phi (k_r) d \bar{c}(k_s) dc (k_t)\exp[i \int d^4 x{\cal L}]}-
\nonumber\\
&&-\langle \phi (k_m)\phi(-k_m)\rangle +2\langle c(k_m)\bar{c}(-k_m)\rangle+\langle A_{\mu}(k_m)A^{\mu}(-k_m)\rangle,
\label{eq3244}
\end{eqnarray}
where the quantities in the brackets are defined similarly with the definition for the Higgs boson.

One can compute the full partition function (in which no additional measure of integration is introduced) to extract the dependence on the momenta \cite{Jora1},
+\cite{Jora2}, \cite{Jora3}, \cite{Jora4}:
\begin{eqnarray}
Z={\rm const}\prod_{n}(\frac{iV}{p_n^2-m_h^2})^{1/2}(\frac{-iV}{p_n^2-m_A^2})^2(\frac{iV}{p_n^2-m_A^2})^{1/2}iV(p_n^2-m_A^2),
\label{res4536}
\end{eqnarray}
where the first factor corresponds to the Higgs boson, the second to the gauge boson, the third to the Goldstone boson and the last one to the ghosts. The Eqs. (\ref{eq3244}) and (\ref{res4536}) lead to:
\begin{eqnarray}
\frac{Vi}{p^2-m_h^2-M(p^2)}=\frac{Vi}{p^2-m_h^2}+\frac{3Vi}{p^2-m_A^2}-\frac{iV}{p^2-m_1^2-\Sigma_1(p^2)}+\frac{2iV}{p^2-m_2^2-\Sigma_2(p^2)}-\frac{4iV}{p^2-m_A^2-\Sigma_A(p^2)}.
\label{res4443566}
\end{eqnarray}
Here all the masses are bare masses and the quantities $M(p^2)$, $\Sigma_A(p^2)$, $\Sigma_1(p^2)$ and $\Sigma_2(p^2)$ are the corrections to the two point functions for the Higgs boson, gauge boson, Goldstone boson and ghost respectively. We can further write Eq. (\ref{res4443566}) as:
\begin{eqnarray}
\frac{1}{p^2-m_h^2}+\frac{3}{p^2-m_A^2}=\frac{1}{p^2-m_h^2-M(p^2)}+\frac{1}{p^2-m_1^2-\Sigma_1(p^2)}-\frac{2}{p^2-m_2^2-\Sigma_2(p^2)}+\frac{3}{p^2-m_A^2-\Sigma_A(p^2)}.
\label{res55345}
\end{eqnarray}
Here in writing the right hand side of Eq. (\ref{res55345}) we took into account two important facts:

1) The actual bare masses $\xi m_A^2$ for the Goldstone boson and the ghost is gauge dependent and both the gauge parameter and the mass can receive quantum correction.

2)  The full propagator in the Feynman  gauge has apart from the correction to the mass the same expression as the bare one since in the model there are no derivative interactions pertaining to the gauge boson.

The full propagator for a boson  particle $X$ has a pole of the form:
\begin{eqnarray}
{\rm Propagator}\approx \frac{1}{p^2-m_X^2-\Sigma_X(p^2)},
\label{res553443}
\end{eqnarray}
where $m_X$ is the bare mass and $\Sigma_X$ is the all order correction to the two point function. It is known that this propagator has a pole at the physical mass of the particle and also weaker singularities in the form of branch cuts.  We expand the denominator in the right hand side of Eq. (\ref{res553443}) to obtain:
\begin{eqnarray}
&&p^2-m_X^2-\Sigma_X(p^2)=
\nonumber\\
&&p^2-m_X^2-\Sigma_X(p^2=m_{Xphys}^2)-\Sigma'(p^2)|_{p^2=m_{Xphys}^2}(p^2-m_{Xphys}^2)-\Sigma''(p^2)|_{p^2=m_{Xphys}^2}(p^2-m_{Xphys}^2)^2...=
\nonumber\\
&&(p^2-m_{Xphys}^2)(1-\Sigma'(p^2)|_{p^2=m_{Xphys}^2}-\Sigma''(p^2)|_{p^2=m_{Xphys}^2}(p^2-m_{Xphys}^2)...)=
\nonumber\\
&&(p^2-m_{Xphys}^2)S_X(p^2)-\Sigma''(p^2)|_{p^2=m_{Xphys}^2}(p^2-m_{Xphys}^2)-...,
\label{res443553}
\end{eqnarray}
where we define:
\begin{eqnarray}
&&m_X^2+\Sigma_X(p^2=m_{Xphys}^2)=m_{Xphys}^2
\nonumber\\
&&(1-\Sigma'(p^2)|_{p^2=m_{Xphys}^2})=S_X=\frac{1}{Z},
\label{def3443}
\end{eqnarray}
and $S_X(m_{Xphy}^2)=\frac{1}{Z}=1$ (the residue of the propagator at the pole is equal  to 1) by the renormalization condition. In consequence:
\begin{eqnarray}
\frac{1}{p^2-m_X^2-\Sigma_X(p^2)}=\frac{Z}{p^2-m_{Xphys}^2}+{\rm \,regular\,\,terms},
\label{form65656}
\end{eqnarray}
and there are no other poles on the right hand side of Eq. (\ref{form65656}) since contributions from two or more particles intermediate states do not produce other poles.

The same expansion can apply to all the quantities on the right hand side of Eq. (\ref{res55345}) and leads to:
\begin{eqnarray}
\frac{1}{p^2-m_h^2}+\frac{3}{p^2-m_A^2}=\frac{1}{(p^2-m_{hphys}^2)}+\frac{1}{(p^2-m_{1phys}^2)}-\frac{2}{(p^2-m_{2phys}^2)}+\frac{4}{(p^2-m_{Aphys}^2)}+ {\rm \,terms\,\, regular}.
\label{rez44298}
\end{eqnarray}
Then by comparing the pole structure on the right hand side and left hand side of the Eq. (\ref{rez44298}) and considering the fact that the quantities $S_A$, $S_1$, $S_2$ have no zeroes we conclude that:
\begin{eqnarray}
&&m_h^2=m_{hphys}^2
\nonumber\\
&&m_A^2=m_{Aphys}^2
\nonumber\\
&&m_{1phys}^2=m_{Aphys}^2=m_A^2
\nonumber\\
&&m_{2phys}^2=m_{Aphys}^2=m_A^2
\label{res553453}
\end{eqnarray}
Note that at first glance there are other possible combinations of results but by considering the possible gauge dependence and the fact that the results should be correct no matter the gauge chosen and also the residue structure for this case it is clear that the possibility outlined in Eq. (\ref{res553453}) is the only choice.  A final remark is in order: these results cannot be generalized easily and may lead to different result for a nonabelian gauge theory due to the presence of the derivative terms in the Lagrangian.

\section{Conclusions}

The renormalizability of the spontaneously broken gauge theories was treated in series of groundbreaking papers \cite{Hooft1}-\cite{Yao}.  The abelian Higgs model has been discussed in detail in \cite{Appelquist}. It is this latter model that we study in the present work. Based on the approach initiated in \cite{Jora1} we analyzed the pole structure of the derivative $\frac{\delta Z}{\delta p^2}$ where $Z$ is the partition function and $p^2$ is the squared momentum, to find that the pole of the Higgs boson all order propagator is situated at the bare mass. The same result is obtained for the massive gauge boson.  However these findings do not imply that the respective propagators do not receive quantum corrections but just that these quantum corrections do not displace the poles from the bare masses. In our study we made the underlying assumption that the the cut-off of the theory goes to $\infty$.

The method employed in this work cannot be extended without complication  to more complex models like the non-abelian gauge theory or the standard model due to the presence of derivative interactions and of the fermion mass terms in these theories. A detailed analysis of these cases will be performed in a separate work.

\section*{Acknowledgments} \vskip -.5cm

The work of R. J. was supported by a grant of the Ministry of National Education, CNCS-UEFISCDI, project number PN-II-ID-PCE-2012-4-0078.

\end{document}